%% file: main.tex
\documentclass[12pt]{iopart}

\usepackage{iopams} 
\usepackage{cite}
\usepackage{slashbox}

\usepackage[utf8]{inputenc} % allow utf-8 input
\usepackage[T1]{fontenc}    % use 8-bit T1 fonts
\usepackage{hyperref}       % hyperlinks
\usepackage{url}            % simple URL typesetting
\usepackage{booktabs}       % professional-quality tables
\usepackage{dsfont}         % stroked numbers
\usepackage{nicefrac}       % compact symbols for 1/2, etc.
\usepackage{microtype}      % microtypography
\usepackage{enumitem}
\usepackage{amsmath}
\usepackage{wrapfig}
\usepackage{graphicx}
\usepackage{cancel}
\usepackage{amsmath}
\usepackage{float}
\usepackage{subcaption}
\usepackage{comment}
\usepackage[absolute]{textpos}

\newcommand\norm[1]{\left\lVert#1\right\rVert}

\usepackage{caption}

\usepackage[dvipsnames]{xcolor}

\def\bea{\begin{eqnarray}}
\def\eea{\end{eqnarray}}

\def\be{\begin{equation}}
\def\ee{\end{equation}}

\makeatletter
\DeclareRobustCommand*{\bfseries}{%
  \not@math@alphabet\bfseries\mathbf
  \fontseries\bfdefault\selectfont
  \boldmath
}

\makeatother

\begin{document}

\title[]{Self-Supervised Learning Strategies for Jet Physics}

\author{Patrick Rieck\textsuperscript{a,\,1}, Kyle Cranmer\textsuperscript{b}, Etienne Dreyer\textsuperscript{c,\,2}, Eilam Gross\textsuperscript{c}, Nilotpal Kakati\textsuperscript{c}, Dmitrii Kobylanskii\textsuperscript{c}, Garrett W.\ Merz\textsuperscript{b}, Nathalie Soybelman\textsuperscript{c}}
\address{\textsuperscript{a} Physics Department, New York University, New York, NY 10003, United States}
\address{\textsuperscript{b} Data Science Institute and Physics Department, University of Wisconsin-Madison, Madison, WI 53706, United States}
\address{\textsuperscript{c} Weizmann Institute of Science, Rehovot, Israel}

\enlargethispage{-1.8\baselineskip}% Shortens the page by 2 lines

\begin{textblock*}{5cm}(1.7cm,26cm)% Adjust coordinates as needed
  \footnotesize $^1$ e-mail: patrick.rieck@nyu.edu\par
                $^2$ e-mail: etienne.dreyer@weizmann.ac.il
\end{textblock*}

\vspace{10pt}
\begin{indented}
\item[]March 2025
\end{indented}
\begin{abstract}
We extend the re-simulation-based self-supervised learning approach to learning representations of hadronic jets in colliders by exploiting the Markov property of the standard simulation chain. Instead of masking, cropping, or other forms of data augmentation, this approach simulates pairs of events where the initial portion of the simulation is shared, but the subsequent stages of the simulation evolve independently. When paired with a contrastive loss function, this naturally leads to representations that capture the physics in the initial stages of the simulation. In particular, we force the hard scattering and parton shower to be shared and let the hadronization and interaction with the detector evolve independently. We then evaluate the utility of these representations on downstream tasks.
\end{abstract}

%%%%%%%%%%%%%%%%%%%%%%%%%%%%%%%%%%%%%%%%%%%%%%%%%%%%%%%%%%%%%%%%%%%%%%%%%%%%%%%
\section{Motivation}

\input{intro.tex}

\section{Related Work}

\input{related_work.tex}

\section{Dataset}

\input{dataset.tex}

\section{Self-Supervised Learning}

\input{ssl.tex}

\section{Results on performance}

\input{results.tex}

%%%%%%%%%%%%%%%%%%%%%%%%%%%%%%%%%%%%%%%%%%%%%%%%%%%%%%%%%%%%%%%%%%%%%%%%%%%%%%%

\section{Results on concepts and learning approaches}

\input{results_concepts.tex}

%%%%%%%%%%%%%%%%%%%%%%%%%%%%%%%%%%%%%%%%%%%%%%%%%%%%%%%%%%%%%%%%%%%%%%%%%%%%%%%
\section{Conclusions} \label{sec:conclusions}

\input{conclusion.tex}

%%%%%%%%%%%%%%%%%%%%%%%%%%%%%%%%%%%%%%%%%%%%%%%%%%%%%%%%%%%%%%%%%%%%%%%%%%%%%%%

%%%%%%%%%%%%%%%%%%%%%%%%%%%%%%%%%%%%%%%%%%%%%%%%%%%%%%%%%%%%%%%%%%%%%%%%%%%%%%%
\begin{ack}

 ED, EG, NK, DK, and NS are supported by the BSF-NSF grant 714179-2020780, the Minerva Grant 715027, and the Weizmann Institute for Artificial Intelligence grant program Ref 151676. PR and KC were supported on NSF grant 2120747. GM is supported by the U.S. Department of Energy (DOE) Award No.~DE-FOA-0002705, KA/OR55/22 (AIHEP). We would like to thank Yann LeCun for insightful conversations that inspired this work.

\end{ack}

%%%%%%%%%%%%%%%%%%%%%%%%%%%%%%%%%%%%%%%%%%%%%%%%%%%%%%%%%%%%%%%%%%%%%%%%%%%%%%%

\section*{References}

\bibliography{main}
\bibliographystyle{naturemag}
% \bibliographystyle{unsrt}

%%%%%%%%%%%%%%%%%%%%%%%%%%%%%%%%%%%%%%%%%%%%%%%%%%%%%%%%%%%%%%%%%%%%%%%%%%%%%%%
\appendix
\newpage

%\section{Appendix} \label{app:appendix}

\end{document}

%% file: intro.tex
The bulk of machine learning in high-energy physics (HEP) is based on a supervised learning paradigm and large simulated datasets where labels are derived from the knowledge of the underlying simulation.  Although this approach is successful for a wide variety of tasks, unsupervised and self-supervised learning paradigms offer promising ways to leverage large, unlabeled datasets and create versatile representations that can be used for multiple downstream tasks.  Self-supervised learning~(SSL) approaches, have seen success in a wide variety of applications including computer vision and natural language processing~\cite{SimCLR}. In science, the SSL approach has proved useful in several fields, including chemistry~\cite{ChemBERTa}, climate science~\cite{climaX}, and astrophysics~\cite{ssl_astro_review}. In HEP, the first explorations of SSL have aimed to derive generic representations of jets that serve as a starting point for various downstream tasks~\cite{JetCLR}. In this paper, we provide new insights into the benefits of SSL for the study of jets, moving this approach closer to applications for particle collider data. In particular, we extend prior work in re-simulation-based self-supervised learning (RS3L)~\cite{Harris:2024sra}. This work moves us closer towards the goal of a foundation model of jet physics.

Most approaches to SSL utilize some notion of data augmentation. Often a single training example is converted into multiple examples that are considered to be intrinsically the same, and the differences between these examples are effectively irrelevant. This notion of 'sameness' may either be an ad hoc heuristic or motivated by some prior knowledge. Examples include randomly cropping an image or transforming a training example according to a symmetry group that is assumed to be present.  In the case of contrastive learning, two data points comprise a ``positive pair'' if they share a common origin and differ only with respect to a randomized augmentation. In the re-simulation-based approach introduced in~\cite{Harris:2024sra}, and extended in this paper, the Markov structure of a stochastic simulator provides the notion of sameness. The loss functions encourage the network to find representations such that positive pairs of data points  are mapped to nearby points in the representation space. Negative pairs of jets, on the other hand, are mapped to points in representation space far apart from each other. In consequence, the resulting neural network, called the backbone or foundation model, which serves as the basis of further unsupervised or supervised downstream tasks, encodes the physics of the jets as it is defined by the underlying notion of sameness. 

The Markov structure of the simulation chain used in HEP, provides a well-motivated and powerful notion of sameness for SSL. The simulation chain is factorized into stages that are based on different phenomena separated by the appropriate energy scales; evolving from high-energy (TeV) to nuclear (GeV), and atomic (MeV) energy scales. Each step in this simulation chain offers a way to define the sameness of a pair of jets, requiring equality of the evolution up to the step of interest, while later steps of the evolution develop independently for the two jets. A first approach along these lines was implemented in~\cite{Harris:2024sra} where the sameness of jets is defined based on the underlying, primary partons. This paper explores an alternative notion of sameness, requiring the sameness of jets down to and including the parton shower step.

In this work, we extend and explore re-simulation-based SSL strategies for HEP and aim to increase similarity of these studies to real physics analyses occurring at collider experiments such as ATLAS and CMS. Our new developments are:

\begin{itemize}

\item A new definition of the sameness of jets underlying the SSL approach, such that the physics of parton showers is incorporated into the learned representation 

\item A realistic detector simulation to study jets emerging from proton-proton collisions. Using state-of-the-art simulations of interactions between particles
and detector material, we take into account the details of the detector response
to jets which make jet physics challenging.

\item An extended phase space spanning the full coverage of the detector and a range of transverse momenta around 100 GeV, which covers the bulk of jets under investigation at CERN’s Large Hadron Collider (LHC)

\item A transformer architecture representing the state-of-the-art models used in HEP experiments~\cite{Qu:2022mxj,ATL-PHYS-PUB-2023-021}

\item Two downstream, supervised learning tasks, namely the classification of quark jets as opposed to gluon jets, and the identification of tau leptons decaying into hadrons 

\item A comparison with a self-supervised pre-training approach other than contrastive learning, namely an autoencoder

\item A demonstration of how to leverage a large dataset by means
of self-supervised learning to detect anomalous jets

\end{itemize}

%% file: related_work.tex
Early work on the applications of SSL for high-energy physics focused on the incorporation of physical symmetries into representations of jets~\cite{JetCLR}.
These representations were derived based on contrastive learning, using the \texttt{NT-Xent} loss function~\cite{SimCLR}, which we also adopt.
In~\cite{JetCLR}, the usefulness of the jet representations based
on SSL was demonstrated in terms of classification tasks, discriminating 
between jets emerging from pure strong interactions on the one hand,
and top-quark jets on the other hand.
In Refs.~\cite{AnomalySSL,DarkCLR}, techniques to identify signatures of physics beyond the Standard Model (SM) of elementary particles were explored, covering a broad range of the new physics parameter space.
In Ref.~\cite{Zhao:2024kry}, large, unlabelled datasets were exploited for means of jet classification.

In Ref.~\cite{Harris:2024sra}, the authors introduced re-simulation-based self-supervised learning (RS3L), which we were independently investigating. In this approach, the pairs of training examples used in the contrastive learning objective are based on a notion of sameness that is motivated by the structure of the simulation chain. In particular, they create pairs of events where the parton shower and detector simulation vary, but the parton-level description of the event is the same. We build on this approach by intervening in the simulation after the parton shower and at the hadronization stage, and we use a more realistic \textsc{Geant4}-based detector simulation. This leads to representations that capture more details of the jet substructure that are useful for a wide range of downstream tasks.

%% file: dataset.tex
\input{jet_sameness.tex}

The dataset used in this study has been created such that it closely resembles the particle detector signals created by jets in the multipurpose experiments at the LHC. We use established tools to create an event simulation workflow that provides pairs of jets suited for SSL, incorporating all common steps of the Markov process that constitutes the simulation of particle collisions and detector interactions as outlined in Fig.~\ref{fig:jet_sameness}.
Each event is seeded by a hard scattering process resulting in two
partons which evolve into jets. We simulate each event twice,
using the same evolution up to and including the parton shower,
but independent evolutions of the hadronisation and
detector interaction.
After the simulation of these pairs of dijet events,
we determine pairs of jets, taking one jet from each of the events
so that we typically obtain two pairs of jets for each pair of events.
We require similar magnitudes and directions of the jet momenta
at the detector as well as particle levels. In addition, we require
the primary partons associated with the two jets to have the same flavour.
As a result, these jet pairs contain the same particle evolution
up to and including the parton shower, which gets encoded by the representations derived from SSL.
The features used for the SSL task are the charged particle tracks and calorimeter energy deposits associated with the jets.

The event simulation aims to generate realistic sets of jets based on proton-proton collisions at a center-of-mass energy of 13\,TeV. For this purpose, we begin the event simulation with the generation of the hard scattering of partons yielding pairs of quarks, gluons, or tau leptons in the final state, using the MadGraph5\_aMC@NLO event generator~\cite{Alwall_2014}. In view of jet classification tasks discussed further below, we choose to generate the quark, gluon, and tau pair final states such that their event yield ratios amount to 3:3:1. The pseudorapidities $\eta$ of the final state particles are limited according to $|\eta|<3.0$, matching the acceptance of the detector simulated downstream. We require each final state particle to have a transverse momentum of $p_{\mathrm{T}}>100$\,GeV, such that the resulting dataset represents the bulk of jets under study at the LHC’s multipurpose experiments. 

The parton-level events resulting from the hard scattering are passed on to Pythia 8.3.10~\cite{Bierlich:2022pfr} for the generation of the parton shower, the formation of hadrons, and the decay of tau leptons.
%A common set of hard scattering events is used and a splitting into pairs suitable for SSL is performed within the Pythia-step as described in the following.
Regarding tau leptons, we take into account all of the decays which have one or three charged hadrons. We simulate all events twice in order to obtain pairs of jets suitable for SSL. The pairs of event simulations are based on the same hard scattering event,
and the parton shower evolves equally up to the first evolution step where a hadron is contained in the event record. From this step forward, the evolution of the event splits, using different random number sequences. This approach results in pairs of jet events where the parton showers associated with the jets are the same while the contained hadrons are different.
Figure~\ref{fig:jet_sameness} illustrates this particular definition of jet
sameness, which we are using in the context of self-supervised,
contrastive learning as discussed in the next section.
Regarding tau lepton decays,
we enforce the same decay channels for each decay within a pair
of events, hence making sure that the tau jets used for SSL have
a reasonable degree of sameness.

Additionally, soft interactions of protons are added to every hard scattering event. These pile-up events are generated with Pythia8. Their interaction vertex is randomly distributed across the beam axis using a Gaussian distribution with a width of 50 mm. The average number of pile-up events per hard scattering event amounts to 150.
%
% ToDo: cutoff is 1 GeV in energy on the particle level, and then 1 GeV in pT in Cocoa
%

The hadron-level events are passed on to the detector simulation package COCOA~\cite{cocoa}, which is based on the 
\textsc{Geant4} toolkit \cite{geant1,geant2,geant3}. Hence we achieve an accurate, state-of-the-art simulation of the detector 
response to the jet events. This level of accuracy is valuable in view of the complex physics of jets. The 
detailed detector simulation allows us to investigate how much of the jet properties can be captured with
 the help of SSL, regarding both global properties like the total jet energy and local, jet 
substructure properties. The COCOA detector simulation includes a parametric emulation of tracking for charged particles, followed by
electromagnetic and hadronic sampling calorimeters comprised of three layers each. Support structures and magnets are also included to simulate energy losses from material upstream of the calorimeter.
To limit the computational needs of the detector simulation, primary
particles are neglected if their transverse momentum is less than 1 GeV.
After the event simulation, jets are formed based on the energy 
distribution in the calorimeters, using the anti-k$_{\mathrm{T}}$ algorithm~\cite{Cacciari:2008gp} with a distance parameter 
of R=0.4 in the space of azimuth angle $\varphi$ and pseudorapidity $\eta$. Charged particle tracks
are associated with the jets, requiring a distance between the track and the jet direction of
$R<0.4$, and a distance to the primary vertex along the beam axis of less than 2 mm.

\begin{figure}[t] % [t] places the figure at the top of the page
    \centering
    \begin{subfigure}[b]{0.45\textwidth}
        \includegraphics[width=\textwidth]{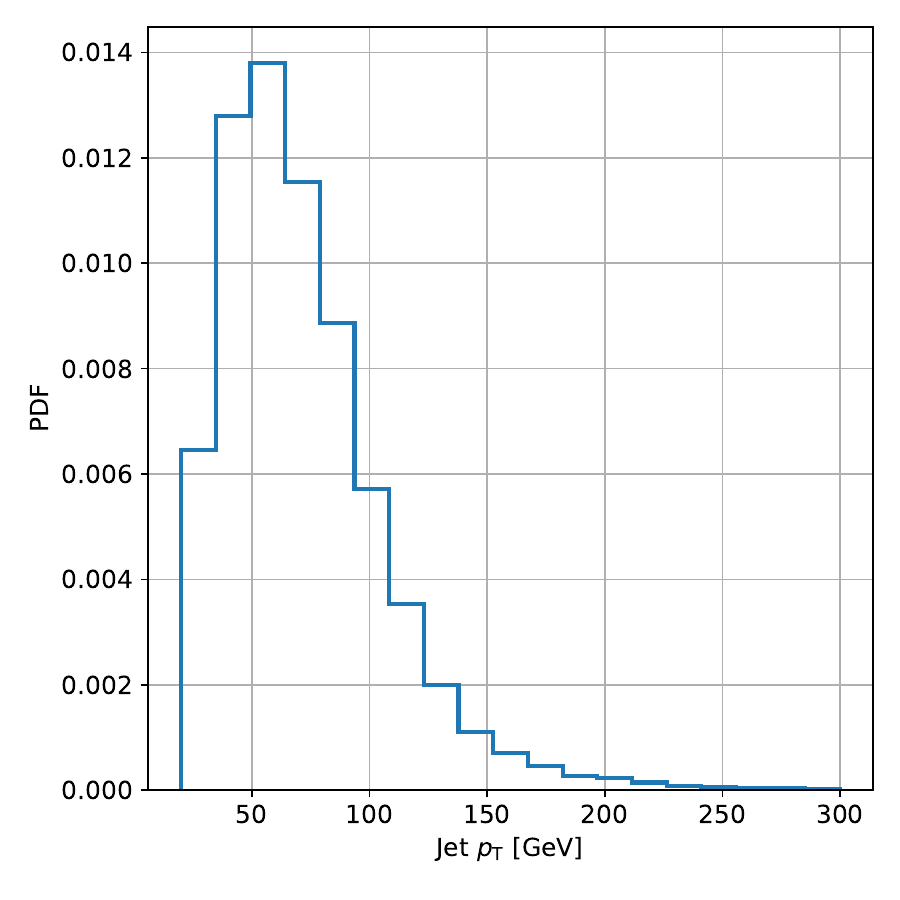}
        \caption{}
        \label{fig:jetPairPt_1d}
    \end{subfigure}
    \quad
    \begin{subfigure}[b]{0.45\textwidth}
        \includegraphics[width=\textwidth]{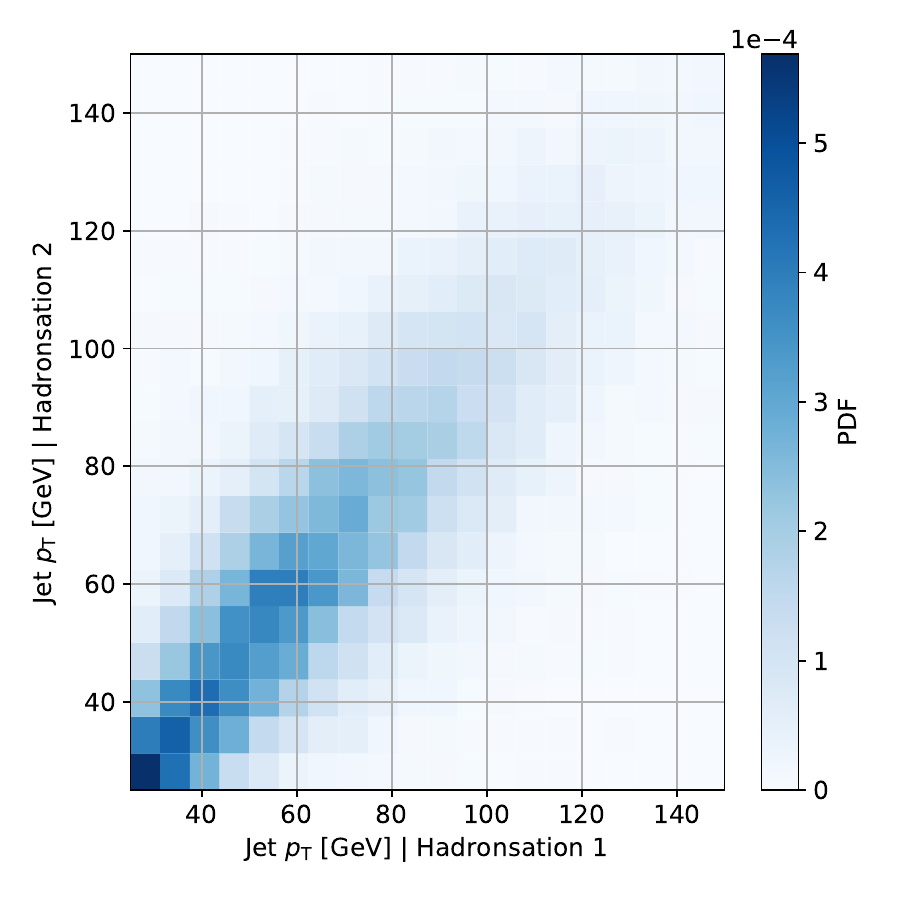}
        \caption{}
        \label{fig:jetPairPt_corr}
    \end{subfigure}
    \caption{Transverse momentum distributions of single jets (\ref{fig:jetPairPt_1d})
    and pairs of jets sharing the same parton shower, used for SSL (\ref{fig:jetPairPt_corr}).
    The jet momenta are not corrected for undetected energy in the calorimeters, leading to
    jet momenta at the detector level which are about a factor of 2 lower than the momenta at the
    hadron level. The correlation among pairs of jets used for SSL amounts to 83 \%.}
    \label{fig:jetPairPt}
\end{figure}

We limit the phase space of jets to be learned using SSL, only jets with a transverse momentum greater than 25 GeV are selected for downstream analysis. Furthermore, for each detector-level jet, the associated jet at the hadron level must have a transverse momentum greater than 50 GeV.
Finally, pairs of jets are created from the two sets of jets distinguished by the random seeds used in the hadronization step and by the detector response. To form a pair, two jets must have the same underlying hard scattering event, the same underlying jet at the hadron level, and a similar momentum at the detector level. Figure~\ref{fig:jetPairPt} presents the distribution of jet transverse momenta. The resulting dataset consists of one million jets, forming half a million jet pairs. Each jet contains an average of 8 charged particle tracks and 800 calorimeter cells.

%% file: jet_sameness.tex
\begin{figure}[t]
    \centering
    \includegraphics[width=0.95\textwidth]{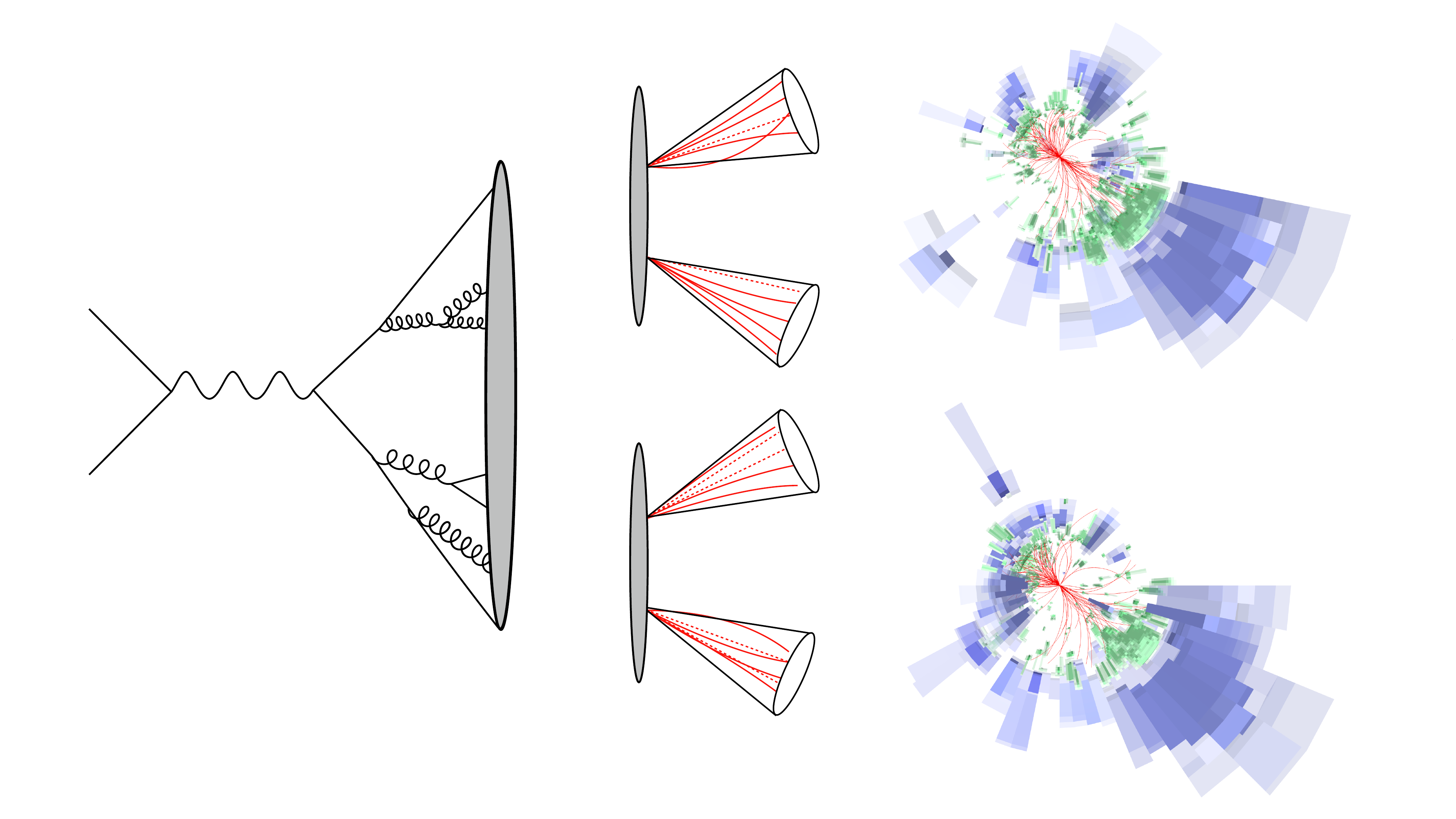}
    \vspace{-8ex} % Space between image and equation

\[
\hspace{-2ex}
\scalebox{3.5}[1]{\rotatebox[origin=c]{90}{$\left\{
\begin{array}{c}
\\
\end{array}
\right.$}}
\hspace{1ex}
\scalebox{2.2}[1]{\rotatebox[origin=c]{90}{$\left\{
\begin{array}{c}
\\
\end{array}
\right.$}}
\hspace{7ex}
\scalebox{2.2}[1]{\rotatebox[origin=c]{90}{$\left\{
\begin{array}{c}
\\
\end{array}
\right.$}}
\hspace{4ex}
\scalebox{6}[1]{\rotatebox[origin=c]{90}{$\left\{
\begin{array}{c}
\\
\end{array}
\right.$}}
\]
    \vspace{-4ex}
\[
\hspace{-8ex}
\begin{array}{ccc}
z_{\mathrm{Hard\,Scatter}} & \hspace{-0.5ex} & z_{\mathrm{Parton}}
\end{array}
\hspace{8ex}
\begin{array}{ccc}
z_{\mathrm{Hadron}}^1 & \hspace{10ex} &  x_{\mathrm{Detector}}^1\\[1ex]
z_{\mathrm{Hadron}}^2 & \hspace{10ex} & x_{\mathrm{Detector}}^2
\end{array}
\]
    \vspace{1ex}
\[
p(x_{\mathrm{Detector}}) = \int \mathrm{d}z_{\mathrm{Hard\,Scatter}}\mathrm{d}z_{\mathrm{Parton}}\mathrm{d}z_{\mathrm{Hadron}}\,p(z_{\mathrm{Hard\,Scatter}}, z_{\mathrm{Parton}}, z_{\mathrm{Hadron}}, x_{\mathrm{Detector}})\ ,
\]
\[
p(z_{\mathrm{Hard\,Scatter}}, z_{\mathrm{Parton}}, z_{\mathrm{Hadron}}, x_{\mathrm{Detector}}) =
\]
\[ 
p(x_{\mathrm{Detector}}|z_{\mathrm{Hadron}})p(z_{\mathrm{Hadron}}|z_{\mathrm{Parton}})p(z_{\mathrm{Parton}}|z_{\mathrm{Hard\,Scatter}})
p(z_{\mathrm{Hard\,Scatter}})
\]

    \vspace{0.5cm} % Space between equation and caption
    \caption{Definition of the sameness of jets used for contrastive learning.
    Particle collision events are simulated as a Markov process,
   organized according to the momentum scale involved in the various interactions. Namely, the hard scattering process seeds the collision event, and is followed by the parton shower, the hadronisation, and the detector response. Each of these steps offers a natural opportunity to define a notion of sameness of two jets.
    By forcing a common hard scattering and parton shower for pairs of jets entering
    the contrastive loss function, we obtain representations that capture
    the most relevant features of the jets for many downstream tasks.}
    \label{fig:jet_sameness}
\end{figure}

%% file: ssl.tex
Our model is based on a foundational backbone neural network which is
pre-trained by means of a contrastive self-supervised learning strategy on positive and negative jet pairs as described in more detail throughout this section.
For downstream learning tasks, this backbone model is extended by a
prediction head network.
During the prediction head training, the backbone model is fine-tuned using low-rank adaption~\cite{lora}.

The dataset is split into disjoint subsets dedicated to the tasks of self-supervised
and downstream supervised learning, using 350\,000 jet pairs and 200\,000 jets, respectively.
Two additional datasets of 50,000 jets each are used for validation purposes during network training,
and to test the network performance.

\label{sec:architecture}

\begin{figure}[t]
    \centering
    \begin{subfigure}[b]{0.9\textwidth}
        \includegraphics[width=\textwidth,height=5.8cm,keepaspectratio]{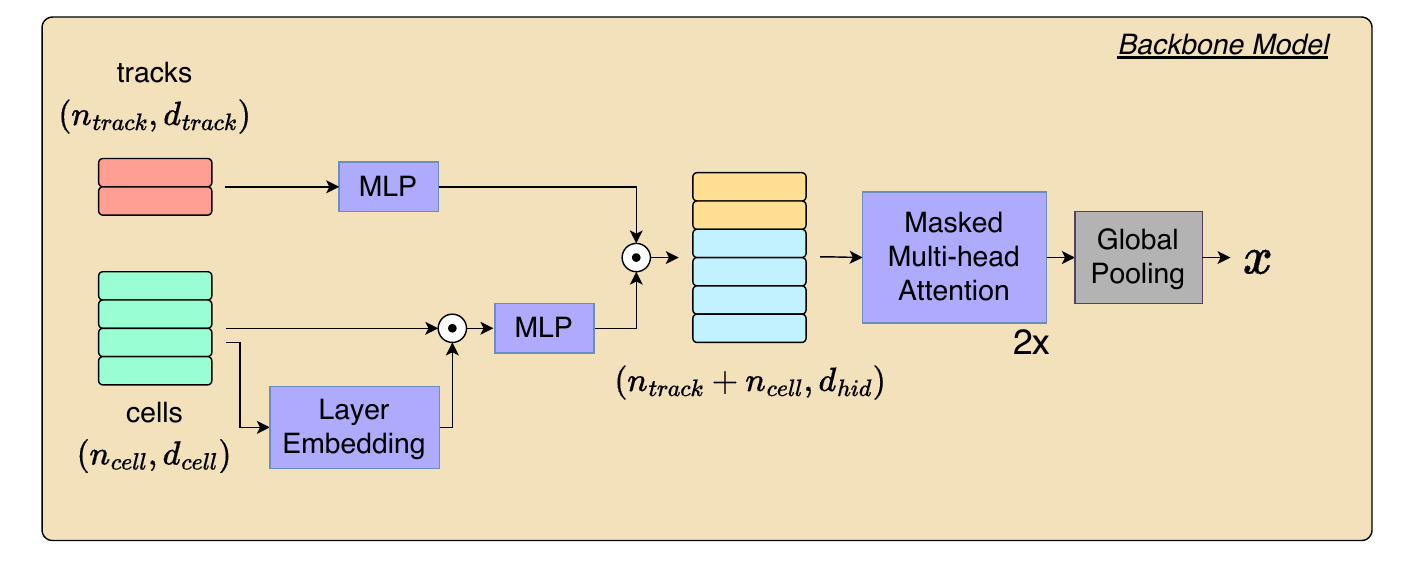}
        \caption{}
        \label{fig:backbone_model}
    \end{subfigure}

    \begin{subfigure}[b]{0.9\textwidth}        \includegraphics[width=\textwidth,height=5.65cm,keepaspectratio]{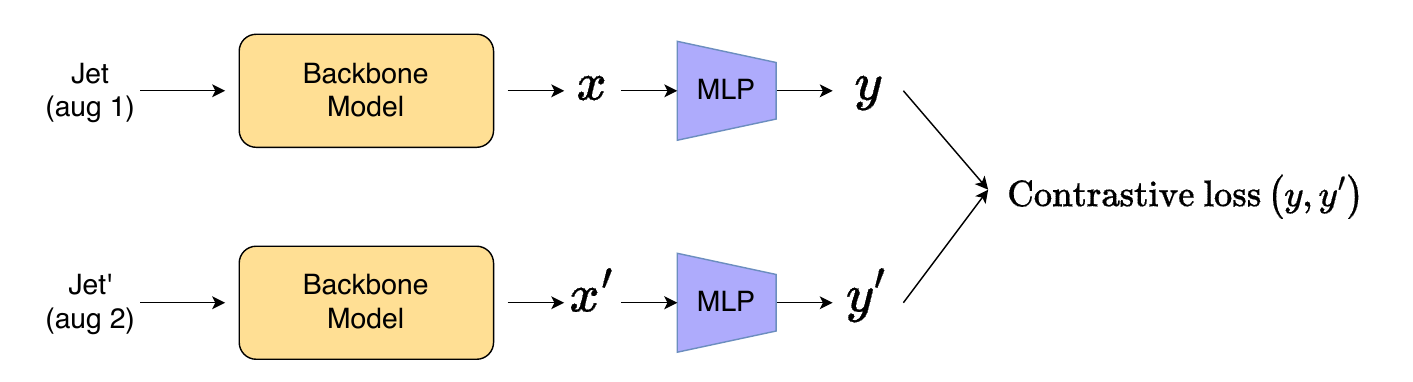}
        \caption{}
        \label{fig:ssl}
    \end{subfigure}
    \caption{SSL foundation model architecture. (a) Charged particle tracks and calorimeter cells associated with a jet are mapped to a same-sized vector space using dedicated MLPs. The resulting representations are updated using a transformer network, and averaged. The $\odot$ sign indicates concatenations. (b) Both jet augmentations are passed through the same backbone model, compressed using another MLP and used for computing the contrastive loss.}
    \label{fig:ssl_backbone}
\end{figure}

The backbone architecture used in the studies is derived from the encoding network used in~\cite{kakati2025hgpflowextendinghypergraphparticle},
where it was used to reconstruct particles using charged particle tracks and calorimeter topoclusters -- aggregated representations of calorimeter cells. In contrast, we use individual calorimeter cells, providing a more detailed input representation that retains information otherwise marginalized within clusters, which we expect to be beneficial.
Figure \ref{fig:ssl_backbone} shows the high-level overview of the architecture chosen for the studies.
The features entering the network are the transverse momentum, pseudorapidity, azimuth angle, as well as the longitudinal and transverse impact parameters of the tracks, and the deposited energy, pseudorapidity, azimuth angle, and layer index of the cells. 
As a pre-processing step, the track transverse momentum, track impact parameters,
and the cell energy are transformed
using a logarithmic function, and all features are normalized to have a mean of zero and a standard deviation of one over the entire training dataset. The detector's calorimeter has six layers - three in the electromagnetic calorimeter (ECAL) and another three in the hadronic calorimeter (HCAL). Since these layers are discrete integer values, the cell layer feature is passed through an embedding layer.

    \begin{table}
        \centering
        \setlength{\tabcolsep}{1.5em} 
        \renewcommand{\arraystretch}{1.2}
        \begin{tabular}{l|r}
            \toprule
            \multicolumn{2}{l}{Hyperparameter} \\ \toprule
            Batch size & 30 \\
            Optimizer & Adam \cite{adam} \\
            Learning rate schedule & Single Cycle~\cite{oneCycleLR} \\
            Maximum Learning rate & 0.001 \\
            Number of epochs & 150 \\ \midrule
            Number of trainable parameters  & \\
            % \; \; \; \; backbone model & 600K \\
            % \; \; \; \; downstream regression model & 20K \\
            % \; \; \; \; downstream classification model & 20K \\
             \; \; \; \; Cell layer Embedding & 24\\
             \; \; \; \; Cell MLP & 17,536 \\
             \; \; \; \; Track MLP & 17,280 \\
             \; \; \; \; Transformer & 199,424 \\
             \; \; \; \; Output Network & 8,776 \\
             \; \; \; \; Downstream MLP & 1,281 \\
            \bottomrule
        \end{tabular}
        \caption{Hyperparameters used in the SSL pre-training and downstream supervised learning.}
        \label{tab:sr_hyperparameters}
    \end{table}

At the first step of the model, the pre-processed track and cell features are mapped to same-sized
vector spaces using two separate multi-layer perceptrons (MLPs). 
After this initialization, the resulting nodes are combined in a single graph. The graph nodes representing the
calorimeter cells and tracks of the jet are connected according
to their degree of proximity, as outlined in detail in~\cite{cocoa}.
The graph representations are updated using a transformer network, using masked multi-head attention~\cite{Vaswani:2017lxt}. The attention masking is based on the edges connecting the cells and
tracks. Subsequently, the transformer output is averaged to obtain a global representation vector $x$.

Another MLP is used to compress the global representation $x$, resulting in an eight-dimensional representation of the jet data, $y$, that enters the %\texttt{NT-Xent}
loss function~\cite{SimCLR}.
Defining positive pairs $(y_i,y_j)\in\mathbb{R}^8\times \mathbb{R}^8$ as two jets with the same underlying parton shower, the contrastive loss is computed as follows:

\begin{equation}
\label{eq:loss}
    \begin{gathered}
    \mathcal{L} = \frac{1}{2N} \sum_{k=1}^N\left[ \ell_{(2k\!-\!1, 2k)} + \ell_{(2k, 2k\!-\!1)}\right],\\
    \ell_{i,j} = -\log \frac{\exp(\mathrm{sim}(y_i, y_j)/\tau)}{\sum_{k=1}^{2N} \mathds{1}_{[k \neq i]}\exp(\mathrm{sim}(y_i, y_k)/\tau)}\ ,
    \end{gathered}
\end{equation}

\vspace{2ex}
\noindent
where the elements of positive pairs $(y_i,y_j)$ are neighbors in the sequence $y_k$, $k\in{1,2,...,N}$, $\mathds{1}_{[k \neq i]} \in \{ 0,  1\}$ is an indicator function evaluating to $1$ if and only if $k \neq i$,
and $\mathrm{sim}(u,v) = u^\top v / \lVert u\rVert \lVert v\rVert$ denotes the cosine similarity between $u$ and $v$.
The one hyperparameter introduced by this loss function is chosen as $\tau=0.1$.
We found stable, optimal results for values of $\tau$ in the range $[0.05, 0.5]$.
During the training procedure, this function forces the network parameters to align the representations of positive pairs in terms of cosine similarity while separating those of negative pairs.

The prediction head networks used in downstream, supervised learning tasks use a hidden layer with 128 nodes to map the normalized,
eight-dimensional representation $y/\lVert y\rVert$ to a single value used for jet classification purposes.

%% file: results.tex
\subsection{Jet Representation Space from Self-Supervised Learning}

First, we investigate the representations $y$ of the jets
determined using SSL as outlined in the previous section,
to evaluate them in view of the contrastive learning aim
and the usefulness for downstream tasks discussed further below.

\begin{figure}[t]
    \centering
    \begin{subfigure}[b]{0.45\textwidth}
        \includegraphics[width=\textwidth,height=5.8cm,keepaspectratio]{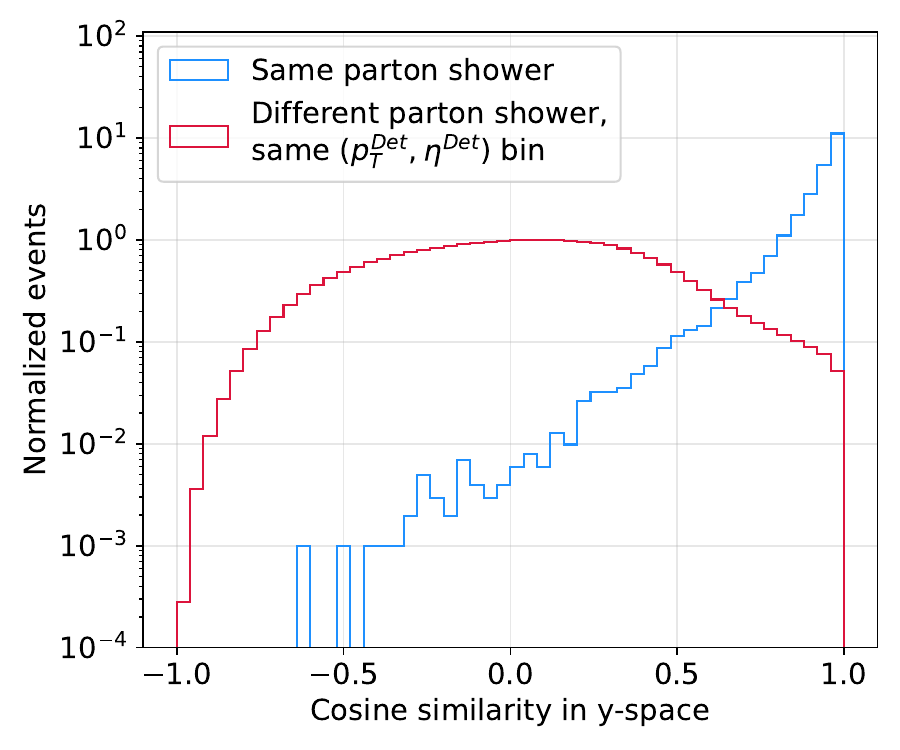}
        \caption{}
        \label{fig:cosine_dist_y}
    \end{subfigure}
    \quad
    \begin{subfigure}[b]{0.45\textwidth}
        \includegraphics[width=\textwidth,height=5.65cm,keepaspectratio]{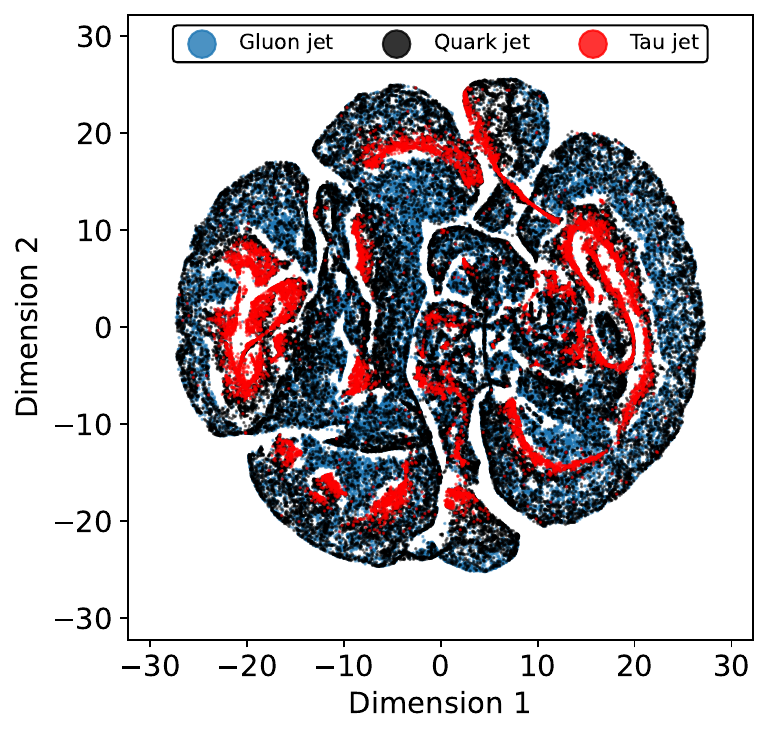}
        \caption{}
        \label{fig:tsne}
    \end{subfigure}
    \caption{Jet representations resulting from SSL.
             Figure~\ref{fig:cosine_dist_y} shows the cosine similarity for
             pairs of jets. To evaluate the SSL of local jet properties, the jets in each pair under investigation are restricted to similar
             rapidity and total energy. Positive pairs result in similar representation vectors, reflecting the SSL of features representing the parton showers of the jets. Figure~\ref{fig:tsne} shows the clustering of jets according
             to their original particle, determined in a two-dimensional projection space used for the purpose of illustration.
    }
    \label{fig:reps}
\end{figure}

Figure~\ref{fig:cosine_dist_y}
presents the cosine-distances $y_1/\norm{y_1}\cdot y_2/\norm{y_2}$ of jet pairs. For positive pairs,
meaning jets with the same underlying parton shower, we obtain cosine distances close to one, indicating
similarity of these jets as required in the SSL. Negative pairs result in lower cosine-distances,
due to the different parton showers of these randomly paired jets. The jets of each pair used in Fig.~\ref{fig:cosine_dist_y}
have similar global properties, namely the rapidity as well as the total energy.
Hence the figure demonstrates the successful SSL of local jet features which are strongly correlated with
the underlying parton showers.
Figure~\ref{fig:tsne} presents the result of a non-linear projection of the jet representation to
a two-dimensional space. Each class of jets, namely quark jets, gluon jets, and tau jets, form
clusters in the representation space.

\begin{figure}
    \centering
        \includegraphics[width=0.9\textwidth]{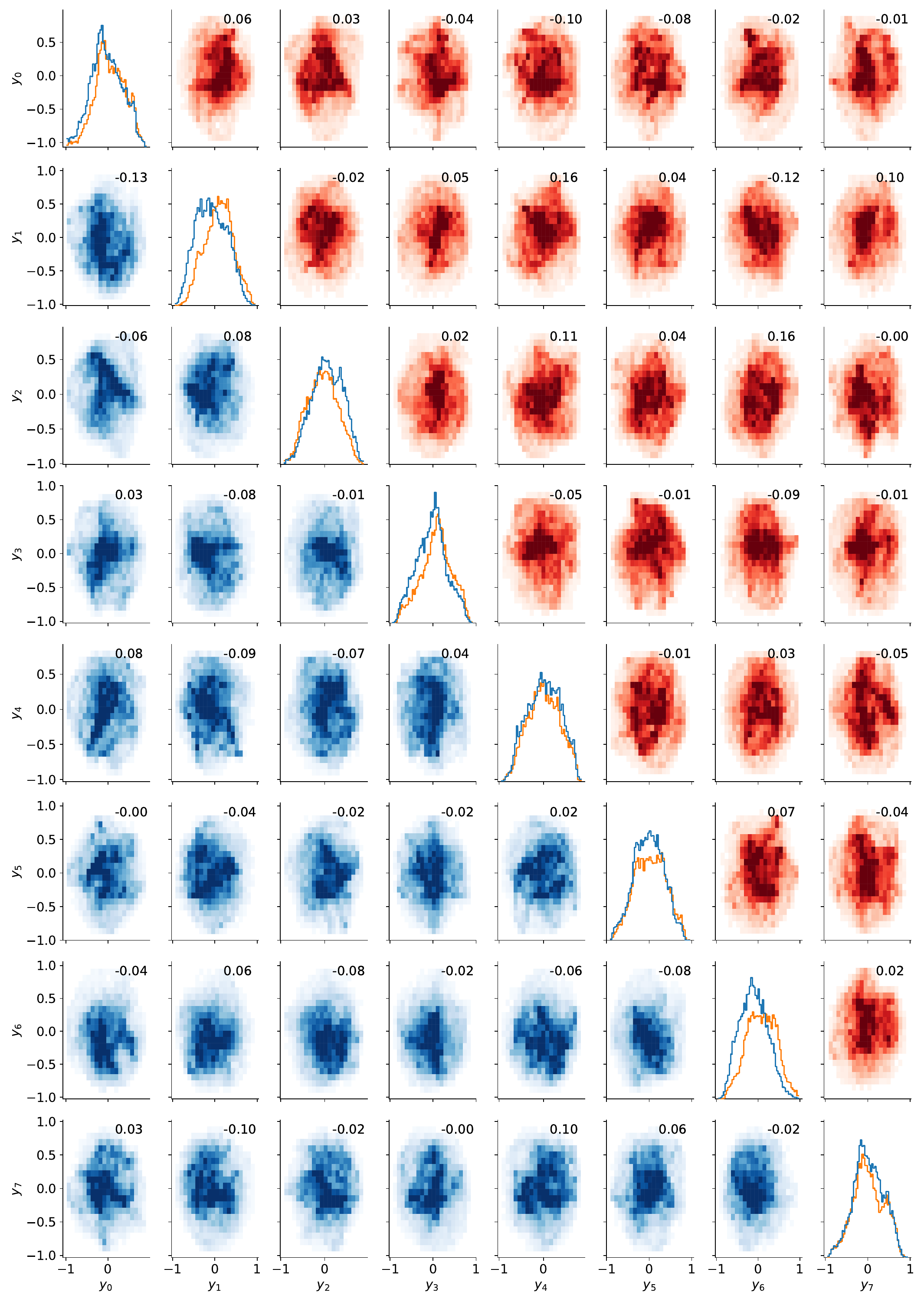}
    \caption{Distributions and correlations of the individual components of the jet representations determined by contrastive learning. The blue figures and curves
    (lower triangle) are quark jets, while the red figures and curves (upper triangle)
    are gluon jets. The one-dimensional distributions are essentialy single peak
    distributions, and all correlations are low, with values on the order of a few percent.}
    \label{fig:corrs_y}
\end{figure}

Figure~\ref{fig:corrs_y} presents the distributions of and correlations among a subset of the 8 components of
the jet representations. The distribution of each component is a simple bell-shaped function with a standard deviation
of the order of one. Furthermore, there are no strong correlations among the different components.

Alternative trainings of SSL backbones with dimensions of y smaller than 8 have been tested, but led to large correlations
among the components of $y$. The prevention of such large correlations using a suitable loss function is left for future work.

\subsection{Clustering of jets due to Self-Supervised Learning}

%
% results for dim_y = 8
%

Based on the representations obtained with SSL, we classified jets using the $k$-nearest neighbour
clustering algorithm. The dataset was split into three disjoint subsets. The first subset was
used to estimate the origin or flavour of a jet in question, defined as the majority of jet 
flavours among the $k$ nearest jets in representation space. Another subset was used to optimize the hyperparameter $k$, yielding similar accuracies for any value $k>10$.
The third subset was used to evaluate the accuracy of the clustering.
This simple approach correctly classifies 71\,\% of gluon jets (comprising 47\,\% of the dataset) and 97\,\% of hadronic taus (11\,\% of the dataset). This clustering of jets of the same flavour demonstrates that
our learning algorithm identifies physical properties of jets without direct supervision.

\subsection{Comparison with fully supervised learning}

\begin{figure}[!t]
    \centering
    \begin{subfigure}[b]{0.45\textwidth}
        \includegraphics[width=\textwidth]{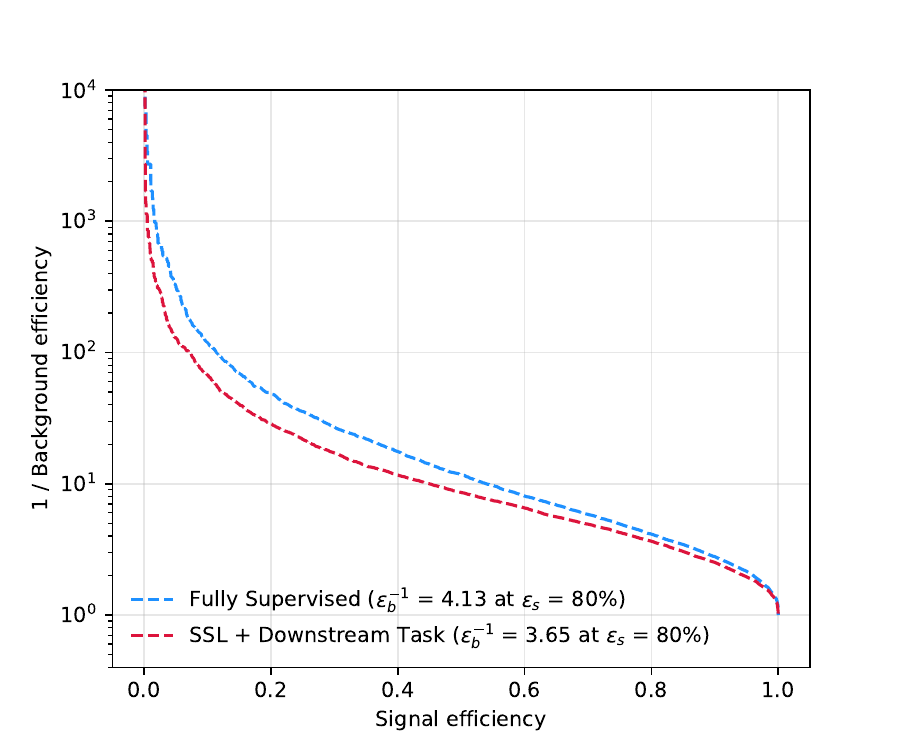}
        \caption{}
        \label{fig:sub3}
    \end{subfigure}
    \qquad
    \begin{subfigure}[b]{0.45\textwidth}
        \includegraphics[width=\textwidth]{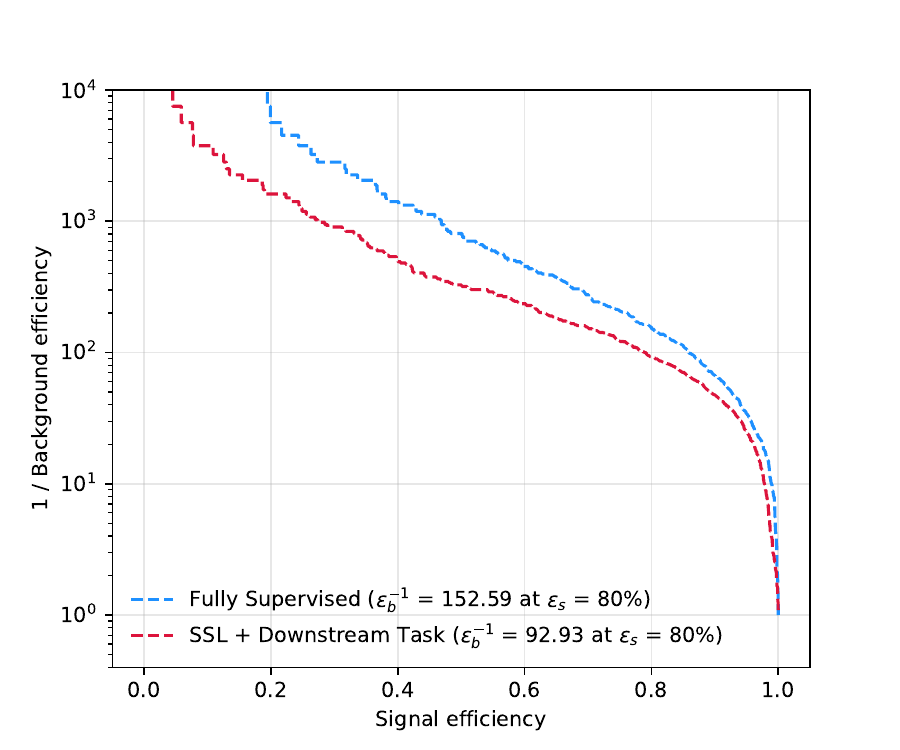}
        \caption{}
        \label{fig:sub4}
    \end{subfigure}\\[1ex]
    \begin{subfigure}[b]{0.42\textwidth}
        \includegraphics[width=\textwidth]{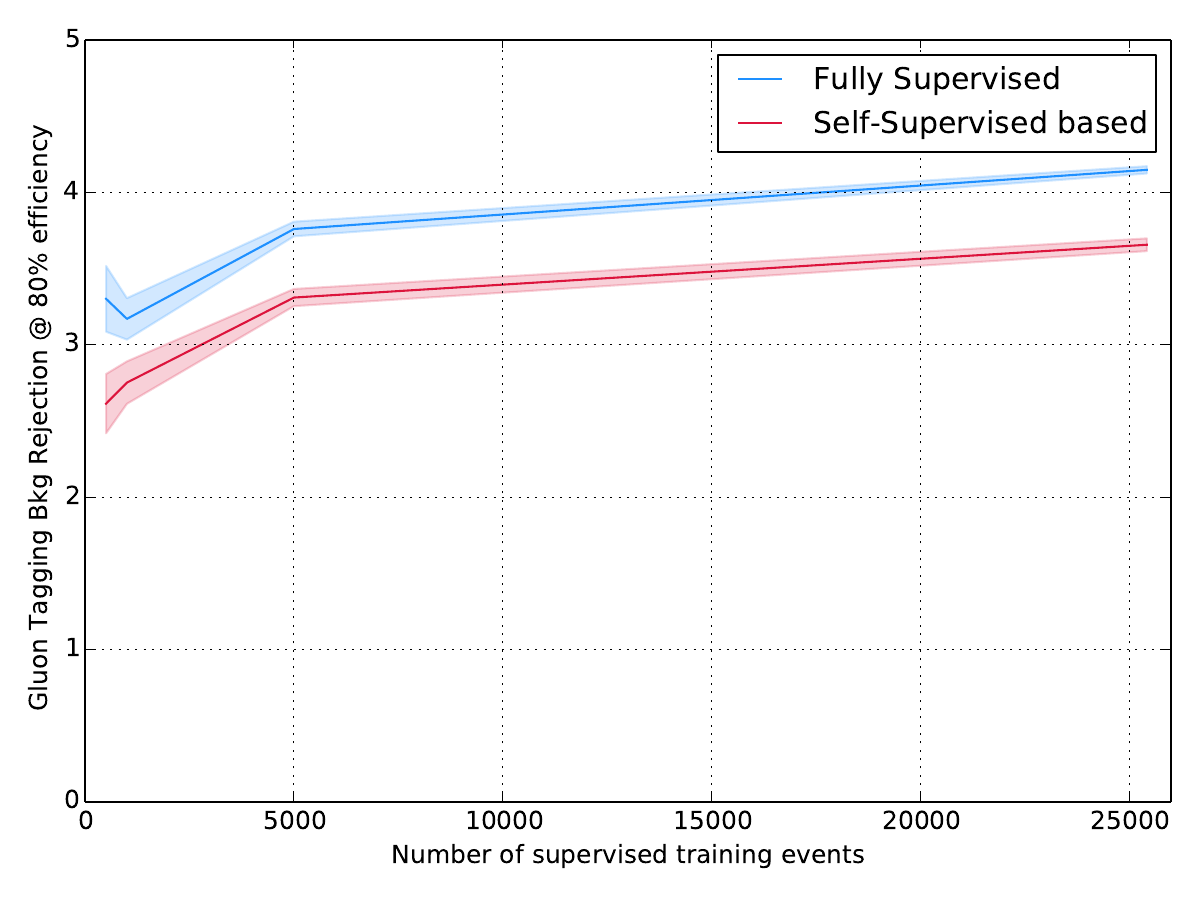}
        \caption{}
        \label{fig:sub5}
        \vspace{-2ex}
    \end{subfigure}
    \qquad
    \begin{subfigure}[b]{0.42\textwidth}
        \includegraphics[width=\textwidth]{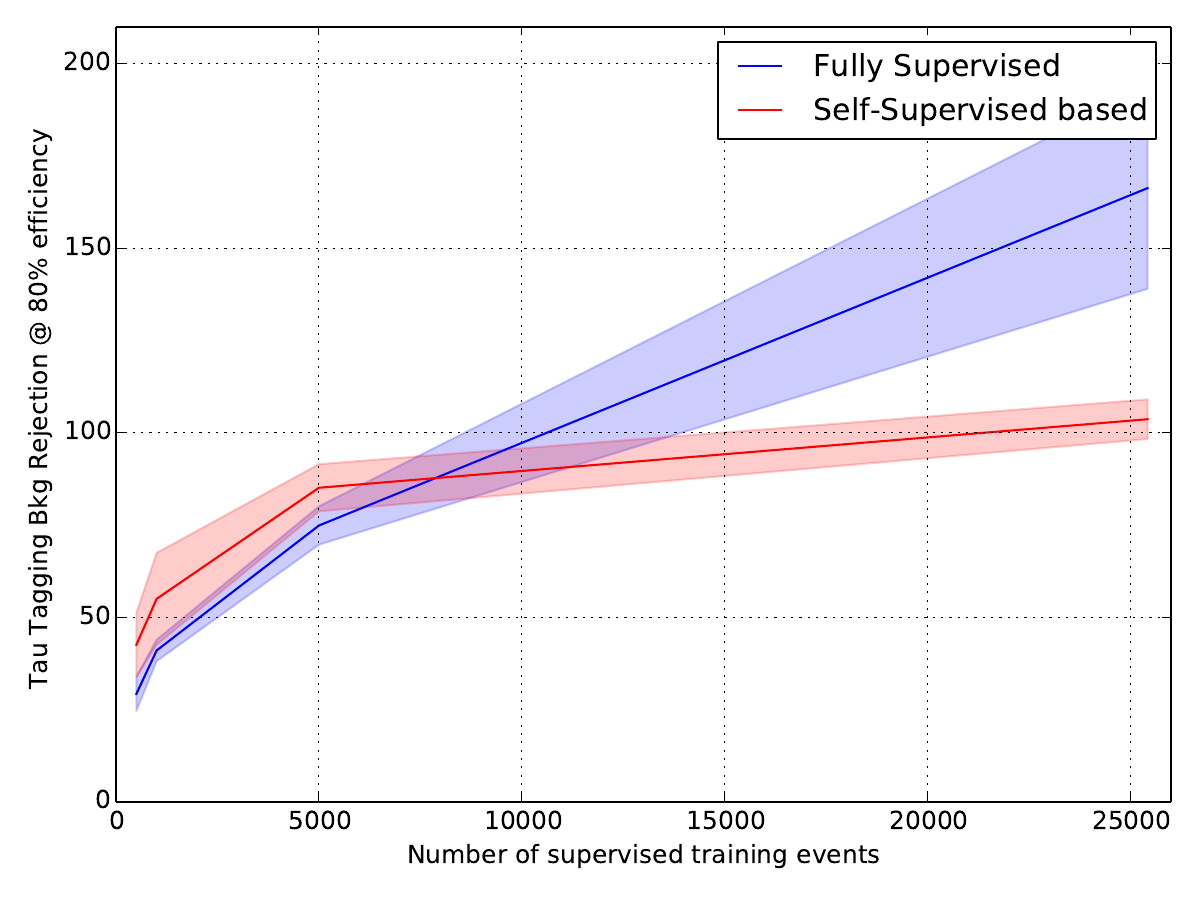}
        \caption{}
        \label{fig:sub6}
        \vspace{-2ex}
    \end{subfigure}
    \caption{Classification of gluon jets (%\ref{fig:sub1}, 
    \ref{fig:sub3}, \ref{fig:sub5})
    and tau jets (%\ref{fig:sub2}, 
    \ref{fig:sub4}, \ref{fig:sub6}), based on self-supervised learning (red curves)
    and fully supervised learning (blue curves).
    %Neural network output distributions are shown in Fig. \ref{fig:sub1} and \ref{fig:sub2}.
    Background rejections and signal efficiencies resulting from cuts on these outputs are shown in Fig.
    \ref{fig:sub3} and \ref{fig:sub4}. The impact of training statistics used in the supervised learning
    step is shown in \ref{fig:sub5} and \ref{fig:sub6}. For tau jet identification and limited
    downstream training statistics, self-supervised learning is beneficial.}
    \label{fig:combined}
\end{figure}

Figure~\ref{fig:combined} compares the performance in the context of two
classification tasks for an SSL and a fully supervised learning approach.
In case of the former, we use the pre-trained transformer backbone model,
fine-tuned by means of low-rank adaption, and combined with a prediction head MLP
trained in a supervised fashion.
In case of the latter, we use a network of the same architecture but determine
its parameters in a fully supervised fashion.
The fully supervised approach is slightly superior,
as expected, while the performance gap between the two approaches depends on the task.
The performance of the supervised approach increases as more labeled data is available for supervised training. For small ammounts of labeled training data, the self-supervised approach is superior in the case of tau tagging.

\subsection{Comparison between contrastive learning and an autoencoder approach}

\begin{figure}[t]
    \centering
    \includegraphics[width=0.8\textwidth]{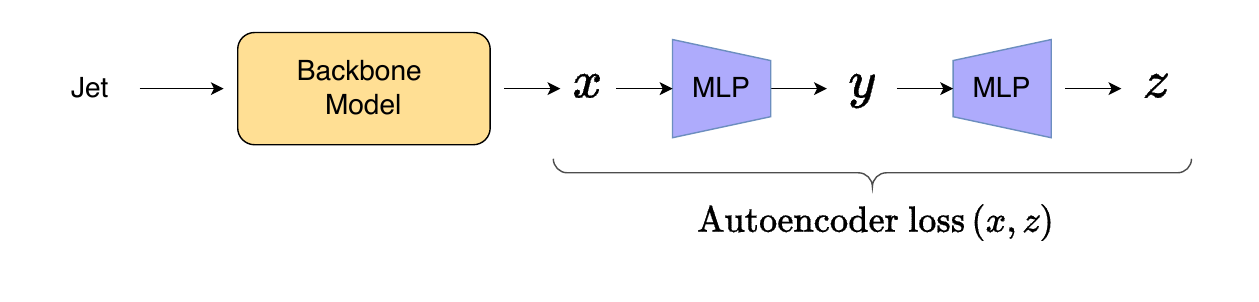}
    \caption{Architecture of the autoencoder approach used as an alternative to contrastive learning. The SSL backbone architecture outlined in Fig.~\ref{fig:backbone_model} is used again. The representation $x$ is compressed to obtain another representation $y$ as in the contrastive learning approach. However, in the autoencoder approach a decompression follows to obtain a representation $z$. The autoencoder training aims to match the representations $x$ and $z$.}
    \label{fig:autoencoder_model}
\end{figure}

To determine the value added by the contrastive learning approach
compared to alternative means of self-supervised learning, we compare it
with an autoencoder approach as outlined in Fig.~\ref{fig:autoencoder_model}.
In this alternative approach, the same backbone is followed by a compressing encoder and decompressing decoder MLP as in a standard
autoencoder. The intermediate compressed form $y$ is used to as the representation of jets.
The loss function used to train this network is the mean-squared error between
the autoencoder input and output. Furthermore, a variance and covariance term
depending on the autoencoder output are added to the loss function to
avoid a collapse of the network that would result only in trivial representations.

We use the compressed representation from this autoencoder model,
denoted $y$ in Fig.~\ref{fig:autoencoder_model}, to perform two supervised learning
tasks where we train a prediction head MLP which classifies its input $y$.
For each of these two learning tasks, we compare the performance of this network with an alternative model that uses the same architecture but stems from a fully supervised training.
In addition, we train a third model where the SSL backbone network
stems from fully supervised learning, but the compressing encoder MLP, which maps the
pooled transformer output $x$ to the target representation $y$, stems from
a dedicated autoencoder training including the decompressing MLP that maps $y$
to $z$.

\begin{figure}[t]
    \centering
    \begin{subfigure}[c]{0.49\textwidth}
        \includegraphics[width=\textwidth]{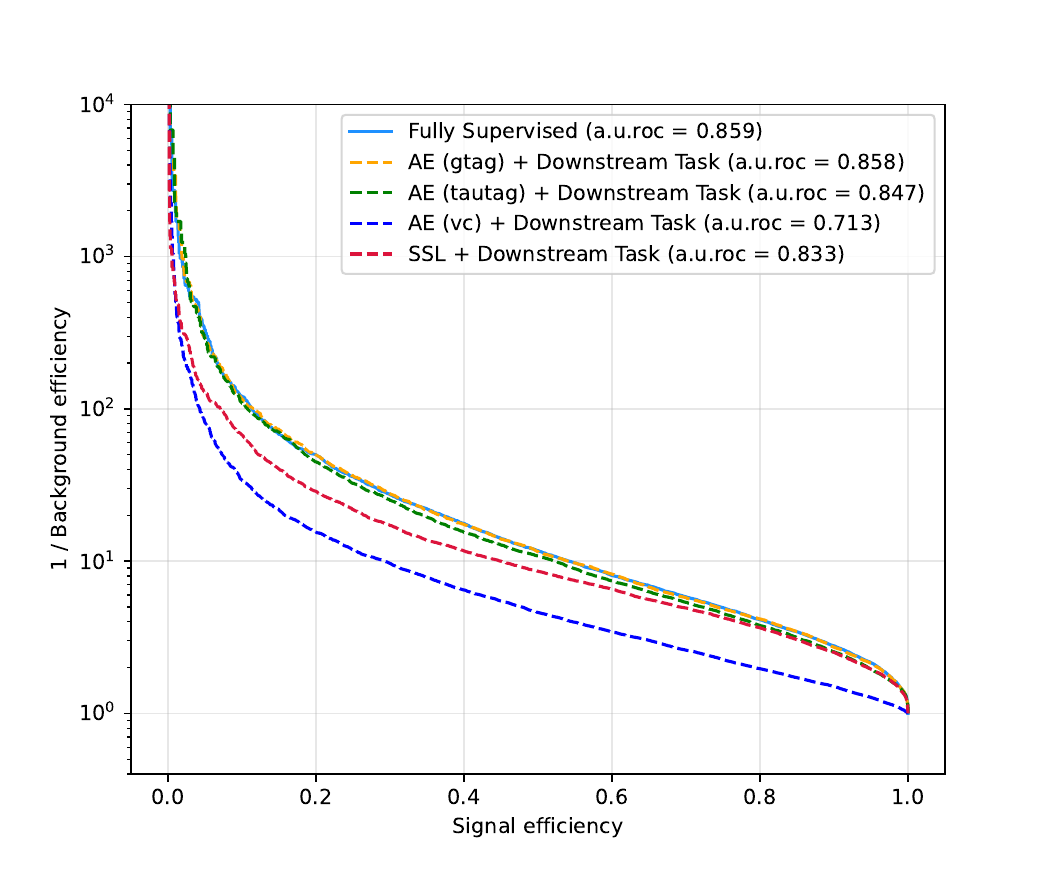}
        \caption{gluon tagging}
        \label{fig:ae_gtag}
    \end{subfigure}
    \begin{subfigure}[c]{0.49\textwidth}
        \includegraphics[width=\textwidth]{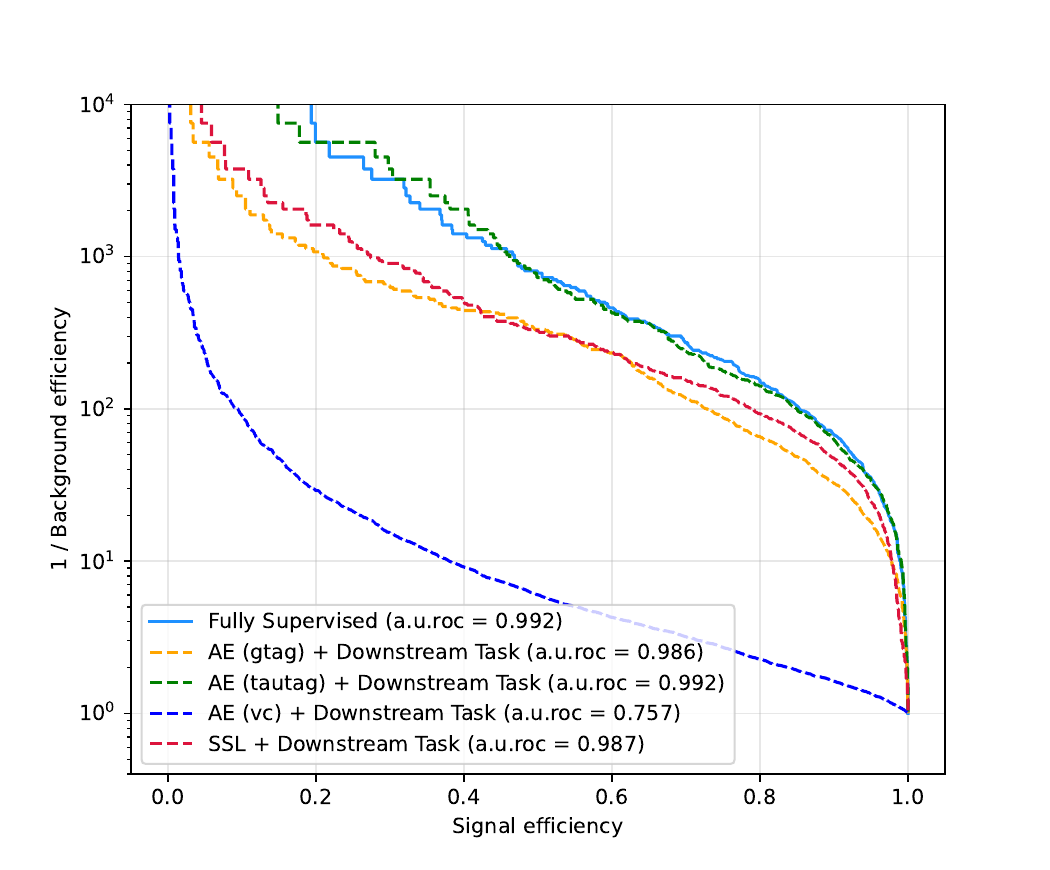}
        \caption{tau tagging}
        \label{fig:ae_tautag}
    \end{subfigure}
    \caption{Performance of an autoencoder approach as an alternative to contrastive learning, investigated for the downstream tasks of gluon tagging and tau tagging.
    Models with autoencoder-based pretraining where the backbone model weights are taken from a supervised training (labelled as AE (gtag) and AE (tautag)) perform as well as models stemming from fully supervised training. However, the backbone network stemming purely from autoencoder training (labelled AE (vc) in the figure), performs worse than the training with a backbone derived from contrastive learning (labelled SSL in the figure).}
    \label{fig:ae_results}
\end{figure}

Figure~\ref{fig:ae_results} shows the resulting background rejection as
a function of the signal efficiency for the two learning tasks, namely
gluon jet tagging and tau jet tagging. The best performance is achieved
by supervised learning as expected. The models combining a pre-trained backbone network
(from supervised learning) with a compressing encoder MLP (from autoencoder training)
perform as good or almost as good as their fully supervised counterparts.
We conclude that the autoencoder training is useful if it is combined
with a pre-trained backbone. On the other hand, the models where both the
the backbone network and the encoder/decoder MLPs of the autoencoder are trained on the unsupervised reconstruction loss objective,
 perform much worse than the other approaches, including
 the model based on pre-training with the SSL contrastive loss objective
(see Fig.~\ref{fig:combined} for comparison with the latter).
Hence we conclude that in our context of learning jet representations,
contrastive learning is the superior self-supervised learning
technique compared to the autoencoder.

\subsection{Application for anomaly detection}

In this section, we build upon the insights presented in the previous section, namely
the value of contrastive learning of a backbone model on the one hand, and the value of the
autoencoder approach used on top of a pre-trained backbone on the other hand.
We combine these two techniques to demonstrate self-supervised learning as a technique to
detect anomalous jets. We build upon the transformer backbone network as outlined previously,
taking advantage of the large training dataset of SM jets and hadronic taus.
We combine this network with an autoencoder as discussed in the previous section,
aiming to minimize the reconstruction loss between the autoencoder's input and output.
We train the autoencoder only on SM jets so that jets of beyond-SM origin will be marked by poor reconstruction.
%Hence the reconstruction loss serves as a measure of jet anomaly when applied to jets of unknown origin.

The backbone network architecture used in this study equals the one used previously.
The autoencoder consists of an encoding MLP made of 5 layers with 128, 64, 32, and 24 nodes,
respectively, a central layer of 20 nodes, and a decoding MLP mirroring the encoding one.
The full SM jet training dataset is used. Unlike the autoencoder in the previous section, we use a different function for the reconstruction loss which is based on the cosine similarity between the input $x$ and output $z$ representations, specifically, $L_\mathrm{reco} = 1-\mathrm{sim}(x,z)$. This choice was motivated by improved sensitivity compared to the mean-squared error loss function.

Testing the capability of this network to detect anomalous jets, we have generated
alternative jet datasets, using so-called dark jet models with four different parameter
sets as used by the ATLAS collaboration~\cite{ATLAS:2023kao}.
These models assume the existence of strongly interacting particles other than the known quarks
and gluons, and that these so-called dark quarks form bound states which are the dark matter
observed in the universe. At particle colliders like the LHC, these dark quarks could be produced
in pairs via weak interactions with the known quarks, and subsequently form jets that contain
both known hadrons and dark, weakly interacting hadrons. This results in anomalous jet signatures.
While using the known phenomenon of strong
interactions to explain dark matter is a well-motivated idea, it suffers from a large space of parameters to explore
in search of such dark jets. Therefore, model-agnostic approaches are valuable for conducting this search,
making dark jets an well-motivated use case for our autoencoder.

We generate jets for each of the four dark jet models,
and resample each dark jet dataset to match the SM
jet distribution in both jet transverse momentum
and pseudorapidity. The resampling rules out the influence of
overall jet momentum differences on the jet classification power.
Hence, this anomalous jet search strategy is restricted to differences in the
substructure of jets, representing scenarios of new physics that are difficult to reveal with
traditional observables.
From the resampled datasets, we take 20\,000 jets for each of the dark jet models.
This is a typical dataset size for a search for new physics at collider experiments,
and it is notably an order of magnitude less data than what we have used in the
pre-training of the transformer backbone network.

\begin{figure}[!t]
    \centering
        \includegraphics[width=0.6\textwidth]{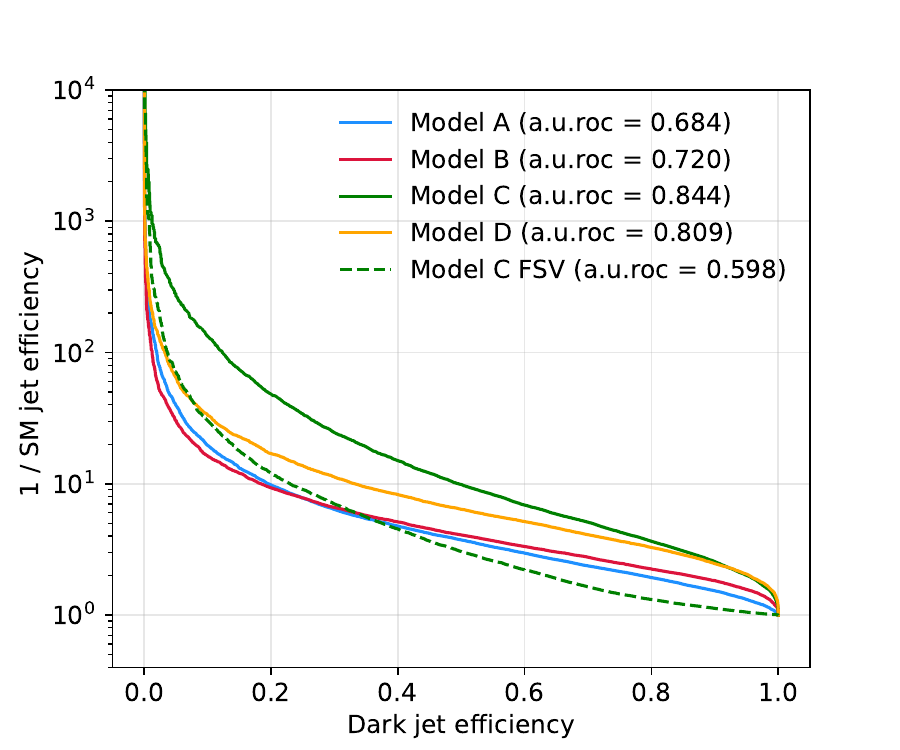}
    \caption{Jet anomaly detection using self-supervised learning.
	A backbone network is pre-trained using contrastive learning
	and combined with an autoencoder. The autoencoder is trained only on
	SM jets using one minus the cosine similarity between input
	and output vectors as the reconstruction loss. Dark jets generated by various new physics scenarios can thus be distinguished by relatively poor reconstruction quality. This approach
	outperforms a fully supervised learning approach (FSV)
	using the same backbone architecture, a prediction head MLP, and a training
	dataset, which is balanced but smaller than the one used in the case of
	self-supervised learning.}
    \label{fig:dj_results}
\end{figure}

We use the logarithm of the reconstruction loss to serve as an anomaly score, quantifying how anomalous the substructure of a jet is. Since we train purely on SM jets, the autoencoder should be able to reconstruct them relatively accurately, with anomaly scores distributing around some negative value. By contrast, the dark jets should exhibit higher (less negative) values of the anomaly score, allowing discrimination between SM and dark jets. We select jets according to various anomaly score thresholds and present the
resulting ROC curves in Fig.~\ref{fig:dj_results}. Clearly, the autoencoder manages to separate dark jets
in each of the four scenarios from SM jets, even though none of the dark jets
were encountered during training of the backbone and autoencoder networks.
We also demonstrate the value of the backbone network pre-training for the dark jet model with the largest anomaly scores (C).
We have trained a network consisting of the aforementioned transformer backbone
and a prediction head MLP in a fully supervised manner to serve as a benchmark classifier. We used a balanced dataset with 60k SM and dark jets each for training and 20k for validation and testing, respectively. 
The resulting performance is shown with a dashed line in Fig.~\ref{fig:dj_results}, where it is clear that the fully-supervised approach is less sensitive than the self-supervised learning one, which benefits from a larger dataset but no anomalous jets. Hence, the necessity
to make assumptions on the nature of the anomalous jets is avoided, and the search is fully
model-agnostic. This result demonstrates how self-supervised learning enables leveraging
large datasets in order to gain physics insights.

%% file: results_concepts.tex
In this section, we report on several findings about the learning schemes
tested throughout this work.

\begin{description}
    \item[Alternative augmentations:] The definition of the sameness of jets used in this work to define pairs entering the contrastive
    learning procedure is one choice among many. Any step in the Markov process that is used to simulate particle collisions  can in principle serve
    to define a notion of sameness of jets. We tested another definition of sameness where we used a simulated hard scattering process
    and ran the rest of the simulation chain twice, starting with the parton shower. Hence the jets considered to form positive pairs
    stem from the same parton in the simulation chain, as in \cite{Harris:2024sra}. Using a dedicated dataset of quark and gluon jets to compare the two
    definitions of sameness, we obtain similar performance for gluon jet tagging. However, we note that our baseline approach
    of including the parton shower simulation step in the definition of jet sameness means incorporating more features in the
    self-supervised learning process, which could be useful for other learning tasks which are not covered in this work and
    left for further research.
    \item[Difference of variance in track and cell input data:] As we are using both charged particle tracks, and calorimeter cells as input data, we needed to take the large difference in variance for these two kinds of input into account. The variance of the tracks is low due to the high granularity of tracking detectors, while the variance is large for calorimeter cells due to the nature of electromagnetic and hadronic showers.
    Hence the contrastive learning procedure can result in representations that neglect the cell input
    and only depend on the track input if it is inherently similar for jets that form a positive pair.
    For instance, this is the case for augmentations that are based on the same particle-level jet
    where only the detector response is varied to give positive pairs of jets. We avoid this failure mode
    by using a different augmentation where jets forming a positive pair have the same parton shower evolution
    but independent, random hadronisation evolutions as well as detector responses as outlined above.
    \item[Particle flow input data:] The low-level detector data provided to the self-supervised or supervised learning
    algorithms can be summarized to varying degrees. While the work presented above is based on a relatively
    low amount of data processing and summarisation before neural network training, namely the use of charged particle tracks as well as calorimeter cell positions and energies as network input, further data processing and summarisation is possible using a particle flow procedure, which summarising track and cell information to form particle candidates as in Ref.~\cite{Harris:2024sra}. We have implemented this alternative procedure using the particle flow algorithm described in \cite{kakati2025hgpflowextendinghypergraphparticle}. In this case, the learning proceeds much faster due to the lower
    number of input quantities, while the performance of the identification of jets stemming from gluons or tau leptons
    slightly drops due to the loss of jet substructure information. However, learning procedures based either on self-supervised or fully supervised approaches still perform similarly, as determined already in the case of the
    track and cell input discussed above.
    \item[Architectures:] As an alternative to the transformer architecture discussed above, a backbone model
    based on a graph neural network has been tested as well~\cite{DiBello2023}, yielding similar qualitative results but overall lower performance. % Ok to make this statement / did we check this for the same number of parameters?
    Furthermore, the foundation model training has also been investigated using an expander network which
    maps the representation $y$ shown in Fig.~\ref{fig:ssl_backbone} to another representation of higher dimension
    as done in Ref.~\cite{SimCLR}. This approach led to a more complicated representation space
    in terms of $y$, with dependencies of the individual components on each other. Therefore, we have not pursued this 
    expander architecture further.
    \item[Fine tuning with low-rank adaption:] The optimisation of the foundation model network using low-rank adaption
    is a valuable compromise between the aims of simplicity of downstream, supervised optimisation tasks on the one hand,
    and their performance on the other hand. For instance, the background rejection of gluon tagging as reported above
    increases by 10\% thanks to this technique used with a rank of 1 for the additional matrices used in the foundation model network, while the number of parameters to be optimised is limited to 1\% of the number of foundation model parameters.
    
\end{description}

%% file: conclusion.tex
We have explored the effectiveness of jet representations obtained from a contrastive learning objective and a physically-motivated notion of sameness that isolates the hard scattering and parton shower. We've compared models that use this common pre-trained backbone for classification and anomaly detection tasks to similar networks that were trained in fully-supervised fashion. We find that the contrastive learning approach provides a common backbone that is permanent for various downstream tasks and that could serve as a foundation model for jets; however, it does not match the performance of a fully supervised model.  We also compare the common backbone trained with contrastive learning to a similar network that has an auto-encoder bottleneck. In that unsupervised approach we use the reconstruction error before and after the encoder-decoder bottleneck as an unsupervised learning objective. We find that the contrastive learning objective outperforms the reconstruction loss objective in the autoencoder setup, and rule out that that this is simply an artifact of the architectural differences by considering a fully supervised version of the same model. 

This work advances our understanding of  re-simulation based SSL strategies with a richer representation that captures the physics of the parton shower. It also exercises these techniques with a more realistic detector simulation with corresponding low-level detector data. SSL continues to be a promising approach that can leverage large unlabled datasets and provide representations that are effective for a number of downstream tasks. 

%%% Comment